**The Solution to the Differential Equation with Linear Damping describing a Physical Systems governed by a Cubic Energy Potential.**

**By:**
   Kim Johannessen, Associate Professor, PhD

   HF & VUC FYN
   Kottesgade 6-8
   5000 Odense C
   Denmark

   Email: kij@vucfyn.dk

**Keywords:**
   Linear damping
   Cubic potential well
   Jacobi elliptic functions
   Period of oscillation
   Asymmetric oscillations
   Perihelion precession

**Abstract**

   An analytical solution to the nonlinear differential equation describing the equation of motion of a particle moving in an unforced physical system with linear damping, governed by a cubic potential well, is presented in terms of the Jacobi elliptic functions. In the attractive region of the potential the system becomes an anharmonic damped oscillator, however with asymmetric displacement. An expression for the period of oscillation is derived, which for a nonlinear damped system is time dependent, and in particular it contains a quartic root of an exponentially decaying term in the denominator. Initially the period is longer as compared to that of a linear oscillator, however gradually it decreases to that of a linear damped oscillator.
   Transforming the undamped nonlinear differential equation into the differential equation describing orbital motion of planets, the perihelion advance of Mercury can be estimated to 42.98 arcseconds/century, close to present day observations of $43.1 \pm 0.5$ arcseconds/century.
   Some familiarity with the Jacobi elliptic functions is required, in particular with respect to the differential behavior of these functions, however, they are standard functions of advanced mathematical computer algebra tools. The expression derived for the solution to the nonlinear physical system, and in particular the expression for the period of oscillation, is useful for an accurate evaluation of experiments in introductory and advanced physics labs, but also of interest for specialists working with nonlinear phenomena governed by the cubic potential well.

**1. Introduction**

   In many interesting areas of physics and engineering the energy potential well governing the motion of the system is conveniently modelled by the quadratic potential of the simple harmonic oscillator, however in addition including a cubic term describing small deviations or anharmonicities of the potential [1,2].

Such systems include, among others the description of the interaction energy of certain diatomic molecules [3,4], where strong repulsive forces dominate at separation distances smaller than the equilibrium distance of the atoms, whereas the anharmonic term takes into account bond breaking of the molecule at larger separation distances, given that the vibrational energy exceeds a certain threshold value. The behavior of certain mechanical oscillators is often modelled by a cubic potential well [5-9], as well as properties of the dynamics of a spherical top, or orbital motion of particles or planets when going from classical Newtonian mechanics to general relativity [10-13]. In the field of marine technology an understanding of the physics of the capsize of vessels can be modelled by the cubic potential [14], or the gravitational collapse of massive stars, or even in cosmology the study of the cubic potential is of interest as a model used in the bosonic string theory [15-17]

Since the energy potential well contains a cubic term, the corresponding differential equation of motion becomes nonlinear with a quadratic nonlinearity. This differential equation also applies for a travelling-wave solution to the Korteweg-de Vries equation [18] or the nonlinear Klein-Gordon equation with quadratic nonlinearity [19]. Numerous studies of the differential equation and its solution have been described, often by numerical methods or by approximate methods such as perturbation techniques [9,20]. However, the exact solution to the undamped and unforced differential equation is well-known in the literature [3,6-8,12,17-19] in terms of the Weierstrassian elliptic functions or using the relation between the Weierstrassian and the Jacobi elliptic functions, it is often equivalently expressed in terms of the more familiar Jacobi elliptic function [12, 20-23].

In the present paper a linear viscous damping term is, however, included in the differential equation and an analytical solution is presented based on the Jacobi elliptic functions, and thus some familiarity with these functions is required. Dealing with elliptic functions one usually has to evaluate an elliptic integral, which often is avoided in introductory physics, since it might be complicated, in addition, it becomes even more complicated when a damping term is included.

Another approach used for solving nonlinear differential equations is to start out by considering the exact solution to the undamped equation and then transfer this solution into a solution including damping. This method has been described by the present author on several occasions [24-26], and has the advantage of being very accurate, including linear damping. This approach has also been applied in the present work to describe the physical system of motion in a cubic potential well.

In the attractive region of the potential well the system becomes an anharmonic and asymmetric damped oscillator. Since the damping term of the differential equation continuously drains out energy of the physical system, the amplitude of oscillation thus decreases with time, and the system eventually becomes a linear harmonic oscillator. This aspect is included in the solution to the differential equation given in terms of the Jacobi elliptic functions, since the elliptic parameter is allowed gradually to decrease to zero, and the elliptic functions thus become equal to the corresponding trigonometric functions.

An expression for the period of oscillation is derived too, which contains a quartic root of an exponentially decaying term in the denominator. Initially the period is longer as compared to that of a linear oscillator, however gradually it decreases to that of a linear oscillator. With the availability of present days accurate timer and sensor systems, it should be possible for students in physics labs to obtain experimental data on the period of oscillation, and compare these data with the theoretical values.

## 2. The cubic potential well

In many physical models the energy potential well governing the motion of the system can be described by the cubic potential well, $V(x) = \frac{1}{2}\alpha x^2 - \frac{1}{3}\varepsilon x^3$, where $x$ is a displacement, $\alpha$ and $\varepsilon$ being some constants in units of energy/length$^2$ and energy/length$^3$, respectively. The first part of the potential describes the quadratic potential of the simple harmonic oscillator, whereas the second part, the cubic term, describes small deviations or anharmonicities of the potential. Due to the cubic term the potential function is not symmetric about the ordinate axis as seen in figure 1, however still it contains an attractive region, when the displacements $x$ is less than $\alpha/\varepsilon$, an unstable equilibrium point at $x = \alpha/\varepsilon$, and an escape region for $\alpha/\varepsilon < x$.

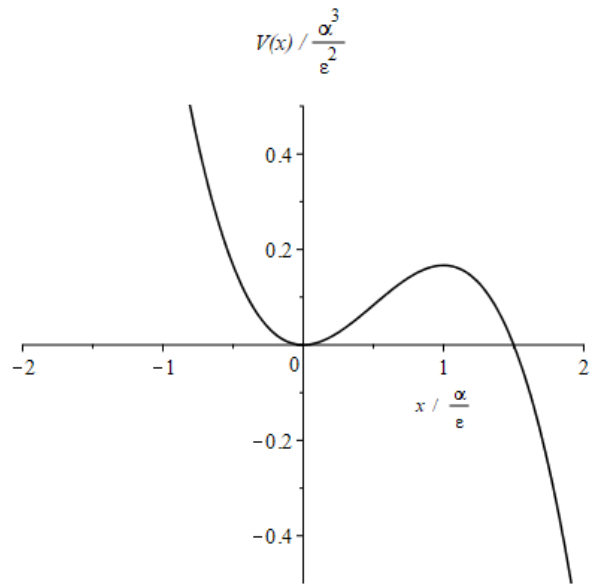

*Figure 1. The cubic potential well, $V(x) = \frac{1}{2}\alpha x^2 - \frac{1}{3}\varepsilon x^3$. The well consists of an attractive region, $x < \alpha/\varepsilon$, an unstable equilibrium point at $x = \alpha/\varepsilon$, and the escape region at $\alpha/\varepsilon < x$.*

Thus a particle trapped in the attractive region of the potential is expected to move with an oscillatory motion, however with asymmetric amplitude since that region is asymmetric, and a particle entering the escape region is expected to move with increasing velocity away from its origin.

The analysis of the physics of a particle moving in a cubic potential well, is divided into the following steps. In section 3 the exact solution to the undamped differential equation is derived. In order to illustrate the usefulness of this solution, in section 4 it is applied to a rather famous physical system, namely the anomalous precession of the perihelion of the planet Mercury. It is not the intend to give a thorough derivation of the equations governing the orbital motion of planets, when going from classical Newtonian mechanics to general relativity, but rather to present a simple way of estimating the precession of the perihelion of planets. In section 5 the exact solution to the undamped physical system is transformed into a solution including linear damping, and in section 6 this solution is compared to the numerical one.

## 3. The solution to the undamped differential equation.

The equation of motion of a particle of mass $m$ moving in an unforced and undamped physical system governed by the cubic potential well, $V(x) = \frac{1}{2}\alpha x^2 - \frac{1}{3}\varepsilon x^3$, is given by:

$$m\ddot{x} + \alpha x - \varepsilon x^2 = 0 \tag{1}$$

The displacement is given by $x$, which is time dependent. In the equation $\alpha$ is a linear stiffness parameter of the harmonic oscillator, and $\varepsilon$ is a parameter describing deviations from the harmonic motion of the system. In order to eliminate the dependence on the mass $m$, equation (1) is rewritten as:

$$\ddot{x} + \alpha_1 x - \varepsilon_1 x^2 = 0 \tag{2}$$

where $\alpha_1 = \alpha/m$ and $\varepsilon_1 = \varepsilon/m$. Since the mass $m$ does not occur explicitly in equation (2), without confusion the letter $m$ will be used in a different meaning in what follows.

The exact solution to equation (2) is well known in the literature, given in terms of the Weierstrassian function $\wp(u)$ [3,6,7,12,17-19], or it can be expressed by the more familiar Jacobi elliptic functions [12,20-23]. In general, the solution procedure of a second order nonlinear differential equation with a quadratic or a cubic nonlinearity is to transform the differential equation into an elliptic integral, and then solve that integral.

However, the approach applied in the present work consists in identifying the indicial equation followed by a transformation of this equation into the appropriate equation of interest. The advantage of this approach is, that only a set of algebraic equations have to be solved, resulting in the necessary mathematical expressions describing the the system.

The indicial equation in the present case related to equation (2) is:

$$\psi'' = 2 - 4(1+m)\psi + 6m\psi^2 \tag{3}$$

where the solution to equation (3) is given by, $\psi(u) = sn^2(u, m)$, $sn(u, m)$ being the Jacobi elliptic function and $m$ now being the elliptic parameter [21,22]. The choice of equation (3) is suggested since the linear and the cubic term of the function $\psi(u)$ come out with the same sign ( when moved to the left hand side of the equation ) as the linear and cubic term of $x$ in equation (2). However, to incorporate the constant term of equation (3), the solution to equation (2) should be expressed as: $x(t) = -a(1 - k\psi(\omega_0 t))$, where $a$, $k$, and $\omega_0$ are constants to be determined. Moreover, this choice of the solution implies, that at time $t = 0\ s$ the initial position and velocity of the physical system is $x(0) = -a$ and $\dot{x}(0) = 0\ m/s$, respectively. For simplicity, these initial conditions are applied in the present work, however different initial conditions would require including a phase shift $\theta$ in the argument of the elliptic function such as $\psi(\omega_0 t - \theta)$. The first and second derivative of $x(t)$ is $\dot{x}(t) = ak\psi'(\omega_0 t)\omega_0$ and $\ddot{x}(t) = ak\psi''(\omega_0 t)\omega_0^2$, respectively. Isolating $\psi''$ and $\psi$ followed by a substitution into equation (3) one finds:

$$\frac{\ddot{x}}{a k \omega_0^2} = 2 - 4(1+m) \frac{x+a}{a k} + 6m \left( \frac{x+a}{a k} \right)^2$$

which after grouping terms leads to:

$$\ddot{x} + x \left( 4(1+m)\omega_0^2 - 12 \frac{m\omega_0^2}{k} \right) - x^2 \frac{6m\omega_0^2}{a k} - 2ak\omega_0^2 + 4(1+m)\omega_0^2 a - \frac{6m\omega_0^2 a}{k} = 0 \qquad (4)$$

Comparison of equation (4) with equation (2) one finds the following system of equations to be solved for the three unknown constants $m$, $k$ and $\omega_0$:

$$4(1+m)\omega_0^2 - 12 \frac{m\omega_0^2}{k} = \alpha_1 \qquad (5.a)$$

$$\frac{6m\omega_0^2}{a k} = \varepsilon_1 \qquad (5.b)$$

$$-2k\omega_0^2 + 4(1+m)\omega_0^2 - \frac{6m}{k}\omega_0^2 = 0 \qquad (5.c)$$

The multiplying factor $\omega_0^2$ of equation (5.c) is kept only to make the algebraic manipulations easier to follow in the next section.

If the parameter $\varepsilon_1$ of the quadratic term of equation (2) becomes vanishing small, i.e. $\varepsilon_1 \to 0$, it simply becomes a harmonic oscillator with the solution $x(t) = -a \cos \sqrt{\alpha_1}\, t$ , given the initial conditions $x(0) = -a$ and $\dot{x}(0) = 0\ m/s$. In that limit it follows from equation (5.b), that the value of the elliptic parameter $m = 0$, and the elliptic functions become identical to the trigonometric ones [21,22]. From equation (5.a) it follows, that $\omega_0 = \sqrt{\alpha_1}/2$ , and from (5.c) that $k = 2$ , thus the solution would be given by $x(t) = -a(1 - 2\psi(\sqrt{\alpha_1}\, t/2))$ , or by doubling the argument, $x(t) = -a(1 - 2\sin^2(\sqrt{\alpha_1}\, t/2)) = -a \cos \sqrt{\alpha_1}\, t$ , in agreement with the result found for the simple harmonic oscillator.

For a parameter $\varepsilon_1$ different from zero the three unknown constants $m$, $k$, and $\omega_0$ can be isolated by use of equations (5.a), (5.b), and (5.c). Substitution of equation (5.b) in equation (5.a) one finds:

$$4(1+m)\omega_0^2 = \alpha_1 + 2\varepsilon_1 a = 4a_1 \qquad (6.a)$$

where a constant $a_1 = (\alpha_1 + 2\varepsilon_1 a)/4$ has been introduced. Substitution of equation (6.a) and (5.b) in equation (5.c) gives:

$$2k\omega_0^2 = \alpha_1 + \varepsilon_1 a \qquad (6.b)$$

However, equation (5.c) can equally well be written ( by use of equations (6.a) and (6.b) ) as:

$$-(\alpha_1 + \varepsilon_1 a) + \alpha_1 + 2\varepsilon_1 a - \frac{6m\omega_0^4}{\frac{1}{2}(\alpha_1 + \varepsilon_1 a)} = 0$$

or

$$m\omega_0^4 = \frac{1}{12}\varepsilon_1 a(\alpha_1 + \varepsilon_1 a) = a_2 \qquad (6.c)$$

where the constant is named $a_2$ for convenience. From equations (6.a) and (6.c) the elliptic parameter $m$ now can be eliminated leading to:

$$4\omega_0^2 + 4\frac{a_2}{\omega_0^2} = 4a_1$$

and thus the three constants $\omega_0$, $k$, and $m$ can be found in succession from the set of equations:

$$\omega_0^2 = \frac{a_1 + \sqrt{a_1^2 - 4a_2}}{2} \qquad (7.a)$$

$$k = \frac{\alpha_1 + \varepsilon_1 a}{2\omega_0^2} \qquad (7.b)$$

$$m = \frac{a_2}{\omega_0^4} \qquad (7.c)$$

where $a_1 = (\alpha_1 + 2\varepsilon_1 a)/4$ and $a_2 = \varepsilon_1 a(\alpha_1 + \varepsilon_1 a)/12$. It should be noted, that the minus sign in the solution of the quadratic equation is discarded, since in the limit of $\varepsilon_1 = 0$ ( where $a_2 = 0$ ) it would lead to $\omega_0^2 = 0$, which is not a valid solution. Moreover, it is seen from equation (7.a) that if $\varepsilon_1 = 0$ then $\omega_0^2 = \alpha_1 / 4$ or $\omega_0 = \sqrt{\alpha_1} / 2$ in agreement with the results previously discussed. Dependent on the sign of the term $a_1^2 - 4a_2$ of equation (7.a), it will either be positive, describing a system in the attractive region of the potential well, zero when the system exactly enters the unstable equilibrium position, and negative for a system entering the escape region in which case either of equations (7.a), (7.b), and (7.c) become complex, however the solution still remains real since the differential equation is real.

The analytical solution ( dashed line ) and the numerical solution ( full line ) to the differential equation (2) is shown in figure 2 for various initial conditions, i.e. $x(0) = -a$, where $a = 0.25\,m$ $a = 0.5\,m$, and $a = 0.75\,m$, and $\dot{x}(0) = 0\,m/s$, and values of $\alpha_1 = 1\,s^{-2}$ and $\varepsilon_1 = 1\,m^{-1}s^{-2}$ have been applied.

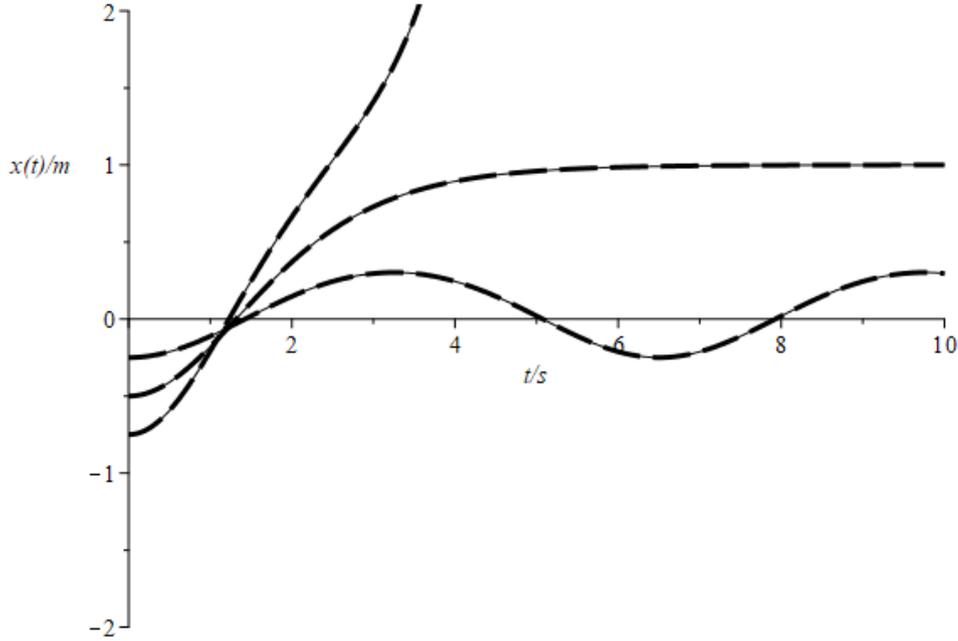

*Figure 2: The analytical solution ( dashed line ) and the numerical solution ( full line ) to the differential equation (2) for various initial conditions, i.e. $x(0) = -a$ , where $a = 0.25\,m$, $a = 0.5\,m$ , $a = 0.75\,m$, and $\dot{x}(0) = 0\,m/s$ . Values of $\alpha_1 = 1\,s^{-2}$ and $\varepsilon_1 = 1\,m^{-1}s^{-2}$ have been applied in this figure.*

It is seen that the analytical and numerical curves coincide and that the oscillatory behavior is asymmetric in the displacement. Although values of $\alpha_1 = 1\,s^{-2}$ and $\varepsilon_1 = 1\,m^{-1}s^{-2}$ have been applied, it should be noted, that the solution $x(t) = -a\,(1 - k\,\psi(\omega_0 t))$, with $\psi(u) = sn^2(u, m)$ , is exact for any combination of $\alpha_1$ and $\varepsilon_1$, positive or negative.

The period of the oscillating solution is found as the following. Since $sn(u, m)$ has a quarterperiod given by [21,22]:

$$K = \int_0^{\frac{\pi}{2}} \frac{1}{\sqrt{1 - m\sin^2(\theta)}} d\theta \qquad (8)$$

the period of $sn(u, m)$ is *4K*, however the period of $\psi(u) = sn^2(u, m)$ is *2K* , since it is squared. Thus the period of the solution, $x(t) = -a\,(1 - k\,\psi(\omega_0 t))$ , is found as $T = 2K/\omega_0$ , giving a value of $T = 6.5004\,s$ ( assuming $a = 0.25\,m$ , $\alpha_1 = 1\,s^{-2}$ and $\varepsilon_1 = 1\,m^{-1}s^{-2}$ ), in agreement with figure 2. In the next section the solution procedure is applied to a very important problem of physics, namely the perihelion precession of the planet Mercury.

## 4. The anomalous perihelion precession of Mercury.

In 1859 it was found by the French astronomer, Le Verrier, that the perihelion of the planet Mercury advanced by an amount of approximately 43 arcseconds per century, an amount that couldn't be explained by the classical Newtonian mechanics. Thus it was clear, that a new physical model was necessary in order to explain the astronomical observation. The General Relativity Theory developed by Einstein was this new model, and in his paper from 1916: "Erklaerung der Perihelbewegung des Merkur aus der allgemeinen Relativitaetstheori", he was able to explain the physics behind the perihelion advance and estimated the advance for Mercury to 43 arcseconds per century, in agreement with observations. Since then, numerous papers and books have been published on General Relativity Theory and the perihelion advance of Mercury. Thus the purpose of this section is not to give a rigorous derivation of the differential equation governing the motion of planets, however instead to apply the procedure of section 3 in solving the differential equation and estimating the perihelion advance of Mercury.

In the equatorial plane ( $\theta = \pi/2$ ) the radial differential equation governing the motion of a planet with the sun at it's origin, is in dimensionless units given by [13] :

$$(u')^2 = u^3 - u^2 + 2\mu u + 2\mu E \qquad (9)$$

In equation (9), $u = u(\varphi) = r_s/r(\varphi)$ , is a dimensionless function dependent on the azimuthal angle $\varphi$ , $r_s$ is the Schwarzschild radius of the sun ( $r_s = 2953.25 m$ ), and $r(\varphi)$ is the radial distance relative to the sun. For the planet Mercury, the initial conditions are such, that at aphelion ( distance furthest away from the sun, where $r_a = 6.98168 \cdot 10^{10} m$ ) the azimuthal angle is $\varphi = 0$ , and the derivative of the radial distance is $r'(0) = 0 \, m$ . The perihelion ( distance of closest approach to the sun, where $r_p = 4.60012 \cdot 10^{10} m$ ) occurs approximately at $\varphi \simeq \pi$ , however not quite, since the planet is precessing around the sun. The quantity $\mu$ is a unit-free orbit parameter related to the angular momentum of the planet, and for Mercury it has approximately the value of $\mu = 5.32497 \cdot 10^{-8}$ [13]. $E$ is a dimensionless total energy, which however, is not needed since we will be considering the derivative of equation (9), which gives:

$$u'' = \mu - u + \frac{3}{2} u^2 \qquad (10)$$

In order to eliminate the constant term on the right hand side of equation (10), the solution is shifted by adding a constant $l$ such that, $u(\varphi) = x(\varphi) + l$ , and thus one obtains:

$$x'' = \mu - (x + l) + \frac{3}{2}(x + l)^2$$

or

$$x'' = \mu - l + \frac{3}{2} l^2 - x(1 - 3l) + \frac{3}{2} x^2 \qquad (11)$$

Chosing $l$ such that the constant term on the right hand side of equation (11) vanishes, one finds that $l = \left(1 + \sqrt{1 - 6\mu}\right)/3$ or $l = \left(1 - \sqrt{1 - 6\mu}\right)/3 = 5.324970576 \cdot 10^{-8}$ , however the first expression is disregarded, since it is expected, that $l << 1$ ( otherwise it wouldn't be consistent with the fact that $u = r_s/r << 1$ ). Moreover, this means that $1 - 3l = \sqrt{1 - 6\mu}$ , and the differential equation now becomes:

$$x'' = -(\sqrt{1 - 6\mu})\, x + \frac{3}{2} x^2 \qquad (12)$$

The differential equations (12) and (2) are similar, except that the time variable $t$ of equation (2) has been replaced by the azimuthal angle $\varphi$ in equation (12). Moreover it is seen that $\alpha_1 = \sqrt{1 - 6\mu} = 0.9999998403$ and $\varepsilon_1 = 3/2$ . From section 3 we thus conclude, that the exact solution to equation (12) is given by $x(\varphi) = -a\,(1 - k\,\psi(\omega_0\varphi))$ , where $\psi(v) = sn^2(v, m)$ , and thus the exact solution to the orbital motion of Mercury is: $u(\varphi) = x(\varphi) + l = l - a\,(1 - k\,\psi(\omega_0\varphi))$ .

It should be noted, that this solution fulfills the criteria, that initially the derivative of the radial distance change at aphelion $r'(0) = 0\,m$ . This follows since $u'(\varphi) = -(r_s/r^2)r'(\varphi)$ and $u'(0) = 0$ , because $\psi'(v) = 2\,sn(v, m)\,cn(v, m)\,dn(v, m)$ where $sn(0, m) = 0$ . Moreover, at the initial azimuthal angle $\varphi = 0$ , we have $u_a = r_s/r_a = 4.229999083 \cdot 10^{-8}$ and, since $\psi(0) = sn^2(0, m) = 0$ , we find the constant $a = l - u_a = 1.094971493 \cdot 10^{-8}$ . From equations (7.a), (7.b), and (7.c) one now finds for the constants $\omega_0$ , $k$ , and $m$ the values $\omega_0 = 0.4999999628$ , $k = 2.000000011$ , and $m = 2.189943324 \cdot 10^{-8}$ , respectively.

The perihelion advance per century for the planet Mercury can now be estimated as:

$$\Delta\varphi = \left(\frac{2K}{\omega_0} - 2\pi\right) 415.2\,\frac{180 \cdot 60 \cdot 60}{\pi} = 42.98''/\,century \qquad (13)$$

where $K = 1.570796335$ is the quarterperiod given by equation (8) of an elliptic function with elliptic parameter $m = 2.189943324 \cdot 10^{-8}$ . The number of revolutions of Mercury per century is 415.2 and the last fraction is the conversion to arcseconds on a semicircle. This estimate of the perihelion advance per century is thus close to present day observations of $43.1 \pm 0.5''$ per century. It should be noted, that since the corrections to the various parameters are relatively small and on the 9'th or 10'th digit ( e.g. $k = 2.000000011$ , where the correction is $0.000000011$ ), the result is very sensitive to rounding off. Thus to avoid rounding off errors a precission of 20 digits has been applied in the calculations.

## 5. The solution to the damped differential equation.

In section 3 a particle of mass $m$ moving in an unforced physical system governed by a cubic potential, $V(x) = \frac{1}{2}\alpha x^2 - \frac{1}{3}\varepsilon x^3$, was analyzed and the exact analytical solution to the differential equation subject to the initial conditions $x(0) = -a$ and $\dot{x}(0) = 0$ was derived. If however linear damping is included the differential equation now becomes:

$$m\ddot{x} + 2\beta \dot{x} + \alpha x - \varepsilon x^2 = 0 \tag{14}$$

The displacement is given by $x$, and $\beta$ is a linear damping coefficient. In analogy with the derivation of the solution to the undamped oscillator the equation is rewritten into the form:

$$\ddot{x} + 2\beta_1 \dot{x} + \alpha_1 x - \varepsilon_1 x^2 = 0 \tag{15}$$

where $\beta_1 = \beta/m$, $\alpha_1 = \alpha/m$, and $\varepsilon_1 = \varepsilon/m$. The mass now is included in the constants and thus in what follows, we will without confusion instead use the letter $m$ to indicate the elliptic parameter. Let's assume that the solution to the damped oscillator is given by:

$$x(t) = -\eta(t) + \varphi(t)\psi(\omega(t)) \tag{16}$$

however in contrast to the undamped physical system, time dependent functions $\eta(t)$, $\varphi(t)$, and $\omega(t)$ now appear in the expression for the solution, which is appropriate since the amplitude decreases due to damping and the frequency becomes nonlinear in time in the initial phase [24-26]. In analogy to the undamped system the function $\psi(u)$ is given by:

$$\psi(u) = sn^2(\xi(u), m(u)) \tag{17}$$

where a scaling function $\xi(u)$ has been introduced. This function is related to the elliptic parameter $m(u)$ by the equation $\xi' - m' f/m = 1$, where $f(u) \cong \frac{1}{2}m(1-m)^{-\frac{3}{2}}(\frac{1}{2} - \frac{3}{16}\frac{m}{1-m})u$, and the specific dependence on the variable $u$ is omitted for simplicity. Details of this approach have been described on several occasions by the present author and can be found in references [24-26]. It enables us in a simple way to incorporate a time dependence of the elliptic parameter, which is necessary, since the parameter is determined by the appropriate amplitude and due to damping, the amplitude is a decreasing function of time in the oscillatory region of the potential well. As a consequence of the relationship $\xi' - m' f/m = 1$, the derivatives of the elliptic functions remain fairly simple, i.e. $dc/du = -s d$, $ds/du = c d$, and $dd/du = -m s d - \frac{1}{2}(m'/m)(1-d^2)/d$, see also references [24-26]. From these relationships for the derivatives it is found, that the differential equation satisfied by equation (17) is:

$$\psi'' = 2 - 4(1+m)\psi + 6m\psi^2 - \frac{1}{2}\frac{m'}{m}\delta\psi' \tag{18}$$

which is similar to equation (3), however now including an extra term due to the time dependence of the elliptic parameter ( see references [24-26] for details).

The function $\delta(u)$ is given by, $\delta(u) = \left(1 - dn^2(\xi(u), m(u))\right)/dn^2(\xi(u), m(u))$, however for $m \to 0$ it follows that $dn(\xi(u), m(u)) \to 1$, thus the last term of equation (18) becomes of less importance for small values of the elliptic parameter. From equation (16), the first and second derivative of $x(t)$ is given by:

$$\dot{x} = -\eta' + \varphi' \psi + \varphi \psi' \omega' \qquad (19)$$

$$\ddot{x} = -\eta'' + \varphi'' \psi + 2\varphi' \psi' \omega' + \varphi \psi'' \omega'^2 + \varphi \psi' \omega'' \qquad (20)$$

where primes denote differentiation with respect to the appropriate variables.

Initially $\omega(0) = 0$ and since $\psi(u)$ is given by equation (17), it follows that $\psi(0) = \psi'(0) = 0$, where $\xi(0) = 0$. From equations (16) and (19) the new initial conditions on the displacement $x(t)$ thus are given by: $x(0) = -\eta(0)$ and $\dot{x}(0) = -\eta'(0)$.

Isolating the derivative $\psi'$ from equation (19) followed by a substitution into equation (20), the latter equation now would only contain terms with $\psi$ and $\psi''$, which in turn can be eliminated by use of equations (18) and (16). After grouping terms the following rather long equation is obtained:

$$\ddot{x} + \dot{x}\left(-2\frac{\varphi'}{\varphi} + \frac{1}{2}\frac{m'}{m}\delta\,\omega' - \frac{\omega''}{\omega'}\right) + x\left(-\frac{\varphi''}{\varphi} + 4(1+m)\omega'^2 + 2\frac{\varphi'^2}{\varphi^2} - \frac{1}{2}\frac{m'}{m}\delta\frac{\varphi'}{\varphi}\omega' + \frac{\omega''}{\omega'}\frac{\varphi'}{\varphi} - 12m\frac{\omega'^2}{\varphi}\eta\right) -$$

$$x^2\frac{6m\omega'^2}{\varphi} + \eta'' - \frac{\varphi''}{\varphi}\eta + 4(1+m)\omega'^2\eta + \left(2\frac{\varphi'}{\varphi} - \frac{1}{2}\frac{m'}{m}\delta\,\omega' + \frac{\omega''}{\omega'}\right)\left(\frac{\varphi'}{\varphi}\eta - \eta'\right) - 2\varphi\omega'^2 - 6m\frac{\omega'^2}{\varphi}\eta^2 = 0$$

$$(21)$$

Since the damping term of the second order differential equation (15) is linear with respect to the velocity, it seems reasonable to assume, that the amplitude is exponentially decreasing, thus the following ansatz is applied $\eta(t) = a\exp(-\beta_1 t)$. Moreover, in order to simplify equation (21) it is further assumed that: $\varphi'\eta/\varphi - \eta' = 0$, from which it follows that: $\varphi'/\varphi = \eta'/\eta = -\beta_1$ and thus: $\varphi''/\varphi = \eta''/\eta = \beta_1^2$. Comparison of equation (21) with equation (15) it is seen, that the terms in front of $\dot{x}$, $x$, and $x^2$ should be equal to $2\beta_1$, $\alpha_1$ and $\varepsilon_1$, respectively. Applying the above mentioned ansatz the following system of equations now has to be solved with respect to the functions: $\varphi(t)$, $\omega(t)$, and $m(\omega(t))$.

$$\frac{1}{2}\frac{m'}{m}\delta\,\omega' - \frac{\omega''}{\omega'} = 0 \qquad (22.a)$$

$$4(1+m)\,\omega'^2 - 12m\frac{\omega'^2}{\varphi}\eta = \alpha_1 - \beta_1^2 \qquad (22.b)$$

$$\frac{6m\omega'^2}{\varphi} = \varepsilon_1 \qquad (22.c)$$

$$4(1+m)\,\omega'^2\eta - 2\varphi\omega'^2 - 6m\frac{\omega'^2}{\varphi}\eta^2 = 0 \qquad (22.d)$$

In the derivation of equation (22.d) it has been used that $\eta'' - \varphi''\eta/\varphi = 0$, which follows from the previously mentioned ansatz. In analogy with the undamped system the multiplying factor $\omega'^2$ of equation (22.d) is kept in order to make the individual steps easier to follow. The content of equation (22.a) is, that as long as the quadratic term of the differential equation is of importance, the elliptic parameter $m(\omega(t))$ as well as the function $\delta(\omega(t))$ both will be different from zero, and thus the function $\omega(t)$ is not linear in time, indicating that the angular frequency of a nonlinear oscillating system is time dependent. If no damping were present, i.e. $\beta_1 = 0$, the solution was found to be $x(t) = -a\,(1-k\,\psi(\omega_0 t))$, thus $\eta(t) = a$, $\varphi(t) = a\,k$, and $\omega(t) = \omega_0 t$, and in that limit equations (22.b), (22.c), and (22.d) are easily seen to be equivalent to equations (5.a), (5.b), and (5.c), respectively. The solution procedure for the functions: $\varphi(t)$, $\omega(t)$, and $m(\omega)$ is similar to that described in the undamped differential equation leading to the following expressions:

$$\omega'(t)^2 = \frac{a_1(t) + \sqrt{a_1(t)^2 - 4a_2(t)}}{2} \qquad (23.a)$$

$$\varphi(t) = \frac{\eta(t)(\alpha_1 - \beta_1^2 + \varepsilon_1\,\eta(t))}{2\omega'(t)^2} \qquad (23.b)$$

$$m(\omega(t)) = \frac{a_2(t)}{\omega'(t)^4} \qquad (23.c)$$

where $a_1(t) = (\alpha_1 - \beta_1^2 + 2\varepsilon_1\,\eta(t))/4$ and $a_2(t) = \varepsilon_1\,\eta(t)(\alpha_1 - \beta_1^2 + \varepsilon_1\,\eta(t))/12$. From equation (23.b) it is seen that the assumption $\varphi'/\varphi = \eta'/\eta$ previously employed is not quite fulfilled, since equation (23.b) consists of a linear and a quadratic term in the function $\eta(t)$, however for small values of the anharmonicity parameter $\varepsilon_1$, it's a reasonable accurate approximation, since the quadratic term vanishes and $\omega'$, as seen from equation (23.a), becomes a constant given by $\omega'(t) = \sqrt{\alpha_1 - \beta_1^2}\Big/2 = b/2$ in that limit. Expressions for the functions $\omega'(t)$, $\varphi(t)$, and $m(\omega(t))$ are given by equations (23.a), (23.b), and (23.c), respectively, and the expression for the function $\omega(t)$, thus can in principle be obtained either by integration of equation (23.a) or more easily by integration of equation (23.c), assuming that the function $m(\omega(t))$ is known. It's the latter approach, that is employed in the following.

Since a scaling function $\xi(u)$ has been introduced related to the elliptic parameter $m(u)$ by $\xi' - m'\,f/m = 1$, an expression for the elliptic parameter as a function of the variable $\omega$, i.e. $m(\omega)$ is needed. In order to obtain an expression for $m(\omega)$ it is necessary to make a number of approximations. From equation (23.a) it follows that as $t \to \infty$ then $\omega'(t)^2 \to (\alpha_1 - \beta_1^2)/4 = b^2/4$, and thus $\omega(t)$ asymptotically becomes $\omega(t) \cong (b/2)t + k_1$, where $k_1$ is an arbitrary constant of integration. From equation (23.c) we will assume, that the elliptic parameter as a function of time approximately is an exponentially decreasing function, which seems reasonable for small values of the anharmonicity parameter $\varepsilon_1$, where the dominant term in $a_2(t) = \varepsilon_1\,\eta(t)(\alpha_1 - \beta_1^2 + \varepsilon_1\,\eta(t))/12$ is exponentially decreasing. The denominator of equation (23.c) consists of the slowly varying function $\omega'(t)^4$, which for large values of time basically is a constant. Thus it is assumed, that $m(\omega(t))$ approximately is given by $m(\omega(t)) \cong (a_2(0)/\omega'(0)^4)\exp(-\beta_1 t) = m_0 \exp(-\beta_1 t)$. The elliptic parameter $m$ as a function of $\omega(t)$ would then be $m(\omega(t)) = m_0 \exp(-\gamma \omega(t))$, where a new constant $\gamma$ has been introduced. However, as $t \to \infty$ we have that $m(\omega(t)) \cong m_0 \exp(-\gamma ((b/2)t + k_1)) \cong m_0 \exp(-\beta_1 t)$ and thus the constant $\gamma$ is approximately given by $\gamma \cong 2\beta_1/b$, neglecting $k_1$ in that limit.

Since $m(\omega(t)) \cong m_0 \exp(-\beta_1 t)$ and $\eta(t) = a\exp(-\beta_1 t)$, an approximate expression for the time dependent function $\omega(t)$ can be derived from equation (23.c) which gives:

$$\omega'(t) = \sqrt[4]{\frac{\frac{1}{12}\varepsilon_1 a}{m_0}\left(b^2 + \varepsilon_1 a \ e^{-\beta_1 t}\right)} = k_2 \sqrt[4]{1 + k_3 e^{-\beta_1 t}} \qquad (24)$$

where the constant $k_2 = \sqrt[4]{\frac{1}{12}\varepsilon_1 a b^2 / m_0} = b/2$, which follows since $\omega'(t) \to b/2$ for $t \to \infty$, and $k_3 = \varepsilon_1 a/b^2$. Thus the following integral now has to be solved:

$$\omega(t) = \frac{b}{2}\int_0^t \sqrt[4]{1 + k_3 e^{-\beta_1 \tau}}\,d\tau$$

The integral can be evaluated by use of the substitution $v^4 = 1 + k_3 \exp(-\beta_1 \tau)$ giving the new limits of the integral: $v_1 = \sqrt[4]{1 + k_3}$ and $v_2 = \sqrt[4]{1 + k_3 \exp(-\beta_1 t)}$, and thus one finds:

$$\omega(t) = \frac{b}{2}\left(\frac{-4}{\beta_1}\right)\cdot\left(v_2 - v_1 + \frac{1}{4}\ln\frac{v_2-1}{v_2+1}\cdot\frac{v_1+1}{v_1-1} - \frac{1}{2}\arctan v_2 + \frac{1}{2}\arctan v_1\right) \qquad (25)$$

which is a function close to linear in time for large values of $t$, however with nonlinear characteristics for small values of time $t$.

Finally, the scaling function $\xi(u)$ which is related to the elliptic parameter $m(u) = m_0 \exp(-\gamma u)$ by the equation $\xi' - m' f/m = 1$ is derived. The specific dependence of these functions on the variable $u$ is omitted for simplicity, and in addition the function $f(u)$ is given by: $f(u) \cong \frac{1}{2}m(1-m)^{-\frac{3}{2}}(\frac{1}{2} - \frac{3}{16}\frac{m}{1-m})u$ ( see references [24-26] for details ). On several occasions the present author has described the derivation of the scaling function [24-26], which is given by:

$$\xi(u) \cong (1 + \frac{1}{4}m(u) + \frac{9}{64}m^2(u))u + \frac{1}{4\gamma}(m(u) - m_0) + \frac{9}{128\gamma}(m^2(u) - m_0^2) \qquad (26)$$

where $\gamma \cong 2\beta_1/b$ and $m_0 = a_2(0)/\omega'(0)^4$. The advantage of introducing such a function is, that a time dependence of the elliptic parameter now is incorporated, however still the derivatives of the elliptic functions remain fairly simple.

In this section the analytical solution to the differential equation describing a quadratic oscillator with linear damping has been derived. Comparison of the analytical solution to the numerical solution found by computer algebra tools ( MAPLE ) is performed in the next section in particular in the attractive region of the potential well, where the oscillatory motion is asymmetric due to the asymmetric potential. An analytical expression for the period of oscillation is presented and found to be very accurate. The motion of the system in the escape region of the potential well is described in the next section too.

## 6. Comparison of the analytical solution to the differential equation for the damped quadratic oscillator with the numerical solution.

An expression for the analytical solution to the differential equation describing a physical system where its motion is governed by a cubic potential well has been derived.

$$\ddot{x} + 2\beta_1\,\dot{x} + \alpha_1\,x - \varepsilon_1\,x^2 = 0 \qquad (15)$$

In this system a linear viscous damping term is included. Since $\psi(0) = 0$ and $\psi'(0) = 0$ it follows from equations (16) and (19) that the initial position and velocity of the system are: $x(0) = -a$ and $\dot{x}(0) = -\eta'(0) = a\,\beta_1$ , respectively, and the solution takes the form:

$$x(t) = -\eta(t) + \varphi(t)\psi(\omega(t)) \qquad (16)$$

where the various expressions have been derived in the previous section.

### The oscillatory region of the cubic potential well.

If the initial displacements $x$ of the physical system is in the interval $-\alpha/2\varepsilon < x < \alpha/\varepsilon$ , the energy level of the cubic potential is $V(x) < \alpha^3/6\varepsilon^2$ , and thus the system is in the oscillatory region of the potential well. Let's assume the initial displacement is $x(0) = -a = -0.1m$, the initial velocity $\dot{x}(0) = a\,\beta_1$ , the linear damping coefficient $\beta_1 = 0.1\,s^{-1}$ and values of $\alpha_1 = 1\,s^{-2}$ , and $\varepsilon_1 = 1\ m^{-1}s^{-2}$ , respectively. The analytical and the numerical solution to the quadratic oscillator is shown in figure 3. The dashed line is the analytical solution given by equation (16), whereas the full line is the numerical solution obtained from equation (15).

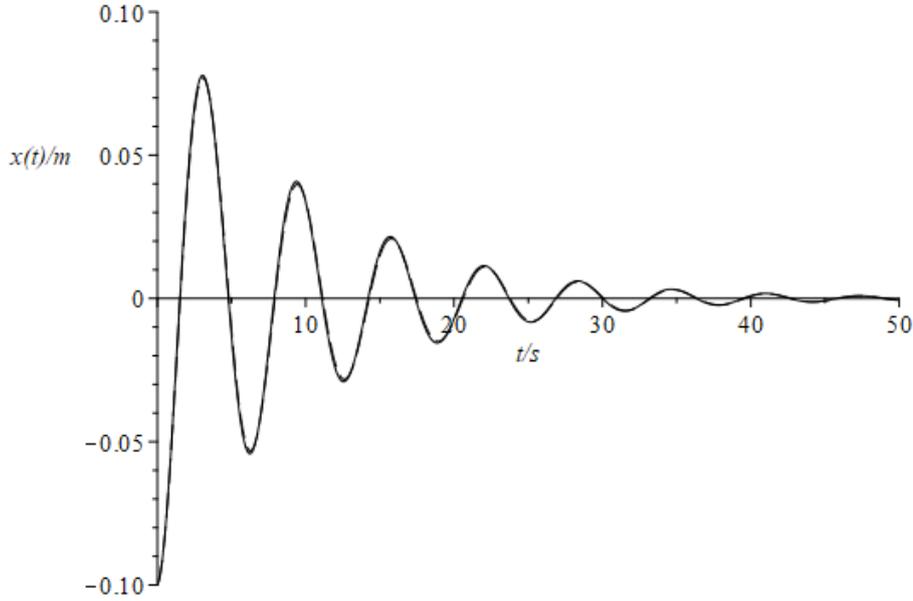

*Figure 3. The oscillatory motion of the quadratic oscillator. The dashed line represents the analytical solution given by equation (16) to the differential equation (15), whereas the full line represents the numerical solution. The value of the damping coefficient is $\beta_1 = 0.1\,s^{-1}$, the values of $\alpha_1 = 1\,s^{-2}$, $\varepsilon_1 = 1\,m^{-1}s^{-2}$, and initial values of $x(0) = -a = -0.1\,m$, $\dot{x}(0) = a\,\beta_1$ have been applied.*

The analytical and the numerical solutions are rather close to one another for these initial conditions. Although one cannot see it clearly in figure 3, due to damping, one should keep in mind, that the oscillatory behavior is asymmetric with respect to amplitude, since the potential well is asymmetric. This behavior was, however, rather pronounced in figure 2 describing the undamped physical system.

The slight discrepancy between these two curves can be ascribed for several reasons. In the derivation of the analytical solution a number of approximations have been applied, one of them being the assumption $\omega''/\omega' \cong 1/2\,(m'/m)\,\delta\,\omega'$ as given by equation (22.a), which is not quite fulfilled. However with increasing time $t$ it becomes more accurate since $\omega(t)$ becomes linear, the elliptic parameter $m(t) \to 0$ and thus $\delta(\omega(t)) \to 0$. In addition, several approximations and truncations of various expressions have been applied in order to obtain a relatively simplified expression for the analytical solution.

The period of oscillation as a function of time for the damped quadratic oscillator can be obtained from the numerical solution in figure 3. However, since the oscillator is asymmetric in amplitude, the procedure applied is, that the time interval for each full cycle ( rising edges or falling edges ) is estimated between the intersection points with the abscissa axis, giving a numerical value of the immediate period, which then can be plotted at the midpoint of the respective interval.

The procedure applied for obtaining an analytical expression for the period of a damped oscillator has been described on several occasions by the present author [24-26]. The main result for the period can be expressed in terms of the quarter period given by equation (8), $K(t)$, which however now is dependent on the elliptic parameter $m(\omega(t))$ of the given Jacobi elliptic function [21,22]. Since the solution is quadratic in the Jacobi elliptic function the period is approximately given by:

$$T^*(t) \cong \frac{2\,K(t)}{\omega^*(t)} \tag{27}$$

where $\omega^*(t)$ is it's immediate angular frequency given by $\omega^*(t) \cong \omega'(t) = \frac{b}{2}\sqrt[4]{1+k_3\,e^{-\beta_1 t}}$. The period of oscillation of the damped quadratic oscillator thus becomes:

$$T^*(t) \cong \frac{4\,K(t)}{b\,\sqrt[4]{1+k_3 e^{-\beta_1 t}}} \tag{28}$$

which is shown as the dashed line in figure 4. The period of oscillation as found from the numerical solution is shown as the dots. Due to the fact, that the oscillatory motion is asymmetric every second dot lies considerably below the theoretical curve. The horizontal line represents the period of the linear oscillator given by $T(t) = 2\pi/b$.

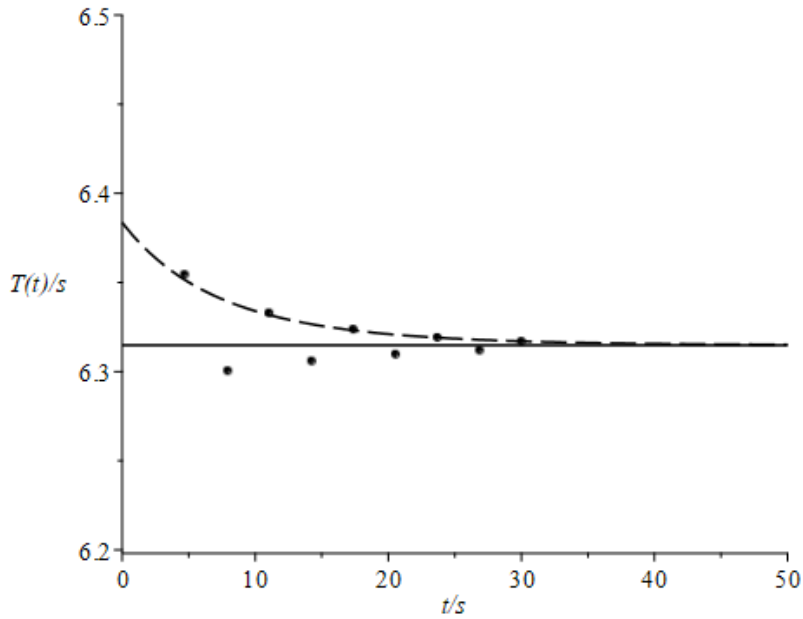

*Figure 4. The period of oscillation of the damped quadratic oscillator. The damping coefficient is $\beta_1 = 0.1\,s^{-1}$ and the values of $\alpha_1 = 1\,s^{-2}$, $\varepsilon_1 = 1\,m^{-1}s^{-2}$ have been applied. The initial conditions are $x(0) = -a = -0.1\,m$ and $\dot{x}(0) = a\,\beta_1$, respectively. Dashed line represents the analytical expression $T^*(t) \cong 4\,K(t)\big/b\,\sqrt[4]{1+k_3 e^{-\beta_1 t}}$, and from the numerical solution the period of oscillation is found and represented as the dots. The horizontal line represents the asymptotic value of the period of oscillation, $T(t) = 2\pi/b$, where $b = \sqrt{\alpha_1 - \beta_1^2}$. Since the oscillatory motion is asymmetric every second dot lies considerably below the theoretical curve.*

It is seen, that every second point of the simulated data are close to the analytic expression for the period of oscillation and the tendency is, that with increasing time the period of oscillation is decreasing, in agreement with results found in various other nonlinear oscillating systems with damping [24-26]. Moreover, it is seen that the expression for the period given by equation (28) for the differential equation with a quadratic anharmonicity contains the quartic root of an expression in the denominator, however if the nonlinearity of the differential equation had been cubic, the expression for the period would be similar, but the root of the denominator would only be quadratic [24-26]. Thus, the analytical expression for the period of oscillation is indicative of the nonlinearity of the differential equation and of the potential governing the motion of the system. A characteristic feature of this plot is, that when the period of oscillation is estimated between the intersections of the rising edges with the abscissa axis in figure 3, one finds values reasonably close to the theoretical expression, whereas if the period is estimated between falling edges, it remains fairly constant close to that of the linear oscillator. This tendency is only observed when damping is included in the differential equation describing an asymmetric potential well.

**The escape region of the cubic potential well.**

If the initial displacements $x$ of the physical system is chosen such that either $x < -\alpha/2\varepsilon$ or $\alpha/\varepsilon < x$, the energy level of the cubic potential is $V(x) > \alpha^3/6\varepsilon^2$, and thus the system enters the escape region of the potential. If the initial displacement is $x(0) = -a = -0.6\,m$, the initial velocity $\dot{x}(0) = a\,\beta_1$, the linear damping coefficient $\beta_1 = 0.01\,s^{-1}$, and values of $\alpha_1 = 1\,s^{-2}$ and $\varepsilon_1 = 1\,m^{-1}s^{-2}$, respectively, the motion of a particle in a cubic potential well would escape to infinity as shown in figure 5.

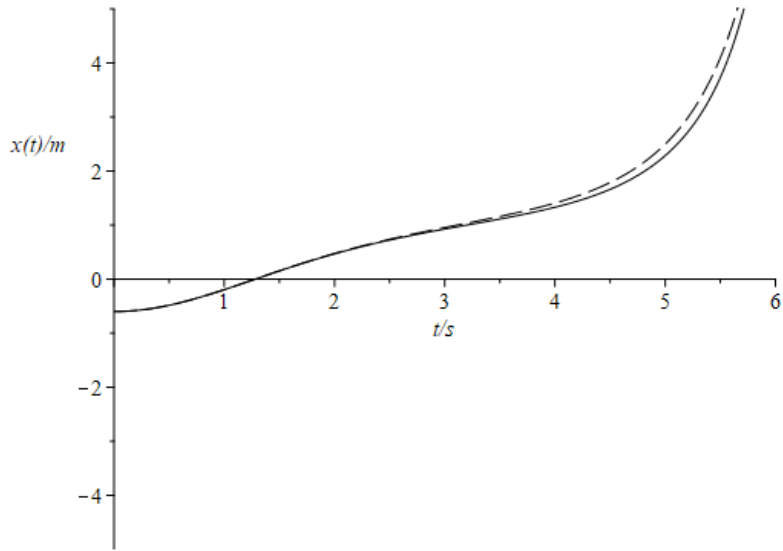

*Figure 5. The escape region. The dashed line represents the analytical solution given by equation (16) to the differential equation (15), whereas the full line represents the numerical solution. The value of the damping coefficient is $\beta_1 = 0.01\,s^{-1}$, the values of $\alpha_1 = 1\,s^{-2}$, $\varepsilon_1 = 1\,m^{-1}s^{-2}$, and initial values of $x(0) = -a = -0.6\,m$, $\dot{x}(0) = a\,\beta_1$ have been applied.*

The dashed line is the analytical solution, whereas the full line is the numerical solution obtained from equation (15). Initially the analytical solution is close to the numerical one, however with increasing time they deviate from one another. This tendency is ascribed to the various approximations applied in the analytical solution, which become of increasing importance in an accurate description of the escape region of the oscillator potential.

# 7. Conclusion

An analytical solution to the nonlinear differential equation, $m\ddot{x} + 2\beta\dot{x} + \alpha x - \varepsilon x^2 = 0$, describing an unforced physical system with linear damping governed by a cubic potential well, is presented in terms of the Jacobi elliptic functions. The elliptic parameter of the Jacobi elliptic functions is time dependent in order to allow for the transition from elliptic to trigonometric functions, which is appropriate since the damping term continuously decreases the displacement of the system and, in the attractive region of the potential well, eventually the quadratic term is only of importance in a perturbative sense. In the attractive region the system becomes an anharmonic damped oscillator, however with asymmetric amplitude because of the cubic term of the potential well.

An expression for the period of oscillation is derived. The expression contains a quartic root in the denominator, which is characteristic for the differential equation with quadratic nonlinearity. Initially the period is longer as compared to that of a linear oscillator, however due to damping it gradually decreases to that of the linear damped oscillator. A significant feature is, that when the period of oscillation is estimated between the intersections of the rising edges with the abscissa axis, one finds values reasonably close to the theoretical expression, whereas if the period is estimated between falling edges, it remains fairly constant close to that of the linear oscillator. This tendency is only observed when damping is included in the differential equation.

The escape region of the cubic potential is described accurately by the solution to the damped nonlinear differential equation, however increasingly deviation between the numerical and analytical solution is observed with increasing time due to the various approximations used in the derivation of the analytical solution.

The solution to the undamped differential equation is exact and applied to the orbital motion of the planet Mercury. The perihelion advance of Mercury is estimated to 42.98 arcseconds/century, close to present day observations of $43.1 \pm 0.5$ arcseconds/century.

The major advantage of the analytical solution to the differential equation describing a quadratic oscillator is, that it describes the attractive and the repulsive region of the potential well accurately, and that it explicitly gives an expression for the period of oscillation. In particular, comparison of experimental data from physics labs. for the period of oscillation with the analytical expression would be very interesting.